\documentclass[showpacs,prl,twocolumn]{revtex4}
\usepackage{amsfonts}
\usepackage{amsmath}
\usepackage{amssymb}
\usepackage{graphicx}

\begin{document}

\title{Hartree-Fock-Bogoliubov Theory of Dipolar Fermi Gases}
\author{Cheng Zhao$^{1,2}$, Lei Jiang$^{1}$, Xunxu Liu$^{1,3}$, W. M. Liu$%
^{3}$, Xubo Zou$^{2}$ and Han Pu$^{1 }$}
\affiliation{$^1$Department of Physics and Astronomy, and Rice Quantum Institute, Rice
University, Houston, TX 77251, USA}
\affiliation{$^2$Key Laboratory of Quantum Information, University of Science and
Technology of China, Chinese Academy of Science, Hefei, Anhui 230026, China}
\affiliation{$^3$Beijing National Laboratory for Condensed Matter Physics, Institute of
Physics, Chinese Academy of Sciences, Beijing 100080, China}

\begin{abstract}
We construct a fully self-consistent Hartree-Fock-Bogoliubov theory that
describes a spinless Fermi gas with long-range interaction. We apply this
theory to a system of uniform dipolar fermionic polar molecules, which has attracted much attention recently, due to rapid
experimental progress in achieving such systems. By calculating the
anisotropic superfluid order parameter, and the critical temperature $T_{c}$, we show that, ``hign $T_c$" superfluid can be achieved with a quite
modest value of interaction strength for polar molecules. In addition, we also show that the presence of the Fock
exchange interaction enhances superfluid pairing.
\end{abstract}

\maketitle

\emph{Introduction. ---} Recent experimental progress in ultracold polar
molecules \cite{jila} has generated great interests in studying the
properties and applications of such system. Applications associated with the
internal energy levels of polar molecules range from quantum information
processing \cite{information} to spin model engineering \cite{spin model}.
An equally intriguing direction is to focus on the external degrees of freedom 
\cite{report}: the system of ultracold polar fermionic molecules with
permanent electric dipoles represents an ideal setup to study dipolar effects in
quantum degenerate fermions \cite{dipolar1,congjun}, as the dipolar interaction strength in these molecular systems is several orders of magnitude larger than that in atomic ones.

Notably, two fundamental properties of the dipolar Fermi gases are superfluid pairing \cite{baranov,bruun} and Fermi surface deformation \cite{taka,sogo}, which are induced by the
partially attractive nature of the dipolar interaction and the anisotropic Fock exchange
interaction, respectively. Mathematically, the long-range interaction
greatly complicates the calculation. As a result, pioneering works such as Ref.~\cite{baranov} and Ref.~\cite{taka,sogo} concentrated on each of these two features and also
made further approximations for simplicity. A quantitatively reliable fully self-consistent theory that includes both these features is lacking.

The goal of the present work is to fill, at least on the mean-field level (which is
believed to be reliable at low temperature for three-dimensional systems), this gap. In order to
achieve this, we construct a self-consistent mean-field theory that takes
full account of the interaction effects. We show how this theory can be efficiently implemented by numerically calculating the superfluid order parameter and the
critical temperature $T_{c}$ for superfluid transition. From our results, we show that robust superfluid
(with $T_c$ being a significant fraction of Fermi temperature) can be
easily reached with ultracold polar molecules. We also investigated the  interplay between Fermi surface deformation and superfluid paring and show that the Fock exchange
interaction enhances superfluid pairing via modifying the
density of states.

\emph{General theory. ---} We consider an ensemble of spinless fermions with a general
two-body interaction potential $U(\mathbf{r},\mathbf{r}^{\prime })=U(\mathbf{%
r}^{\prime },\mathbf{r})$ confined in an
external trapping potential $V(\mathbf{r})$. The second quantized Hamitonian
reads 
\begin{eqnarray}
H &=&\int d\mathbf{r}\,\psi ^{\dag }(\mathbf{r})\left[ -\frac{\hbar
^{2}\triangledown ^{2}}{2m}-\mu +V(\mathbf{r})\right] \psi (\mathbf{r}) 
\notag \\
&+&\frac{1}{2}\int d\mathbf{r}\int d\mathbf{r}^{\prime }\psi ^{\dag }(%
\mathbf{r})\psi ^{\dag }(\mathbf{r}^{\prime })U(\mathbf{r},\mathbf{r}%
^{\prime })\psi (\mathbf{r}^{\prime })\psi (\mathbf{r})\,,  \label{H}
\end{eqnarray}%
where $\psi $ is the fermion field operator, and $\mu $ is the chemical
potential. Denoting $\{\eta (\mathbf{r})\}$ as a complete set of
single-particle eigenstates of $-\hbar ^{2}\nabla ^{2}/(2m)+V(\mathbf{r})$
with eigenenergies $\varepsilon _{\eta }^{0}$, and the associated
annihilation operator $C_{\eta }$, Hamiltonian (\ref{H}) can be rewritten as 
\begin{equation*}
H=\underset{\eta }{\sum }\varepsilon _{\eta }C_{\eta }^{\dag }C_{\eta }+%
\frac{1}{2}\underset{\eta _{1},\eta _{2},\eta _{3},\eta _{4}}{\sum }U_{\eta
_{1},\eta _{2},\eta _{3},\eta _{4}}C_{\eta _{1}}^{\dag }C_{\eta _{2}}^{\dag
}C_{\eta _{3}}C_{\eta _{4}}\,,
\end{equation*}%
Where $\varepsilon _{\eta }\equiv \varepsilon _{\eta }^{0}-\mu $ and 
\begin{equation*}
U_{\eta _{1},\eta _{2},\eta _{3},\eta _{4}}=\int d\mathbf{r}\int d\mathbf{r}%
^{\prime }\eta _{1}^{\ast }(\mathbf{r})\eta _{2}^{\ast }(\mathbf{r}^{\prime
})U(\mathbf{r},\mathbf{r}^{\prime })\eta _{3}(\mathbf{r}^{\prime })\eta _{4}(%
\mathbf{r})\,.
\end{equation*}%
Performing the mean-field decoupling to the quartic operators, we obtain the
effective mean-field Hamiltonian 
\begin{eqnarray*}
&&H_{\mathrm{eff}}=\underset{\eta }{\sum }\varepsilon _{\eta }^{0}C_{\eta
}^{\dag }C_{\eta }+\underset{\eta ,\eta ^{\prime }}{\sum }\{[U_{h}(\eta
,\eta ^{\prime })+U_{f}(\eta ,\eta ^{\prime })]C_{\eta }^{\dag }C_{\eta
^{\prime }} \\
&\,\,-&\!\!\!\!\frac{1}{2}[U_{h}(\eta ,\eta ^{\prime })+U_{f}(\eta ,\eta
^{\prime })]\langle C_{\eta }^{\dag }C_{\eta ^{\prime }}\rangle -\frac{1}{2}%
\Delta (\eta ,\eta ^{\prime })\langle C_{\eta }^{\dag }C_{\eta ^{\prime
}}^{\dag }\rangle \\
&\,\,+&\!\!\!\!\frac{1}{2}\Delta (\eta ,\eta ^{\prime })C_{\eta }^{\dag
}C_{\eta ^{\prime }}^{\dag }+\frac{1}{2}\Delta ^{\ast }(\eta ,\eta ^{\prime
})C_{\eta ^{\prime }}C_{\eta }\}\,,
\end{eqnarray*}%
where the Hartree term $U_{h}$, the Fock term $U_{f}$ and the pairing term $%
\Delta $ are defined as 
\begin{eqnarray*}
U_{h}(\eta ,\eta ^{\prime }) &=&\underset{\eta _{1},\eta _{2}}{\sum }\langle
\eta _{1},\eta |U|\eta ^{\prime },\eta _{2}\rangle \,\langle C_{\eta
_{1}}^{\dag }C_{\eta _{2}}\rangle \,, \\
U_{f}(\eta ,\eta ^{\prime }) &=&-\underset{\eta _{1},\eta _{2}}{\sum }%
\langle \eta _{1},\eta |U|\eta _{2},\eta ^{\prime }\rangle \,\langle C_{\eta
_{1}}^{\dag }C_{\eta _{2}}\rangle \,, \\
\Delta (\eta ,\eta ^{\prime }) &=&\underset{\eta _{1},\eta _{2}}{\sum }%
\langle \eta ,\eta ^{\prime }|U|\eta _{1},\eta _{2}\rangle \,\langle C_{\eta
_{1}}C_{\eta _{2}}\rangle \,.
\end{eqnarray*}
$H_{\mathrm{eff}}$ has a quadratic form and can therefore be diagonalized
using the standard Bogoliubov transformation.

\emph{Polar Fermi Molecules. ---} We now apply the general theory outlined
above to a system of uniform dipolar Fermi molecules with dipole moment $%
\mathbf{d}=d\hat{z}$ polarized along the $z$-axis. It is convenient to study
this problem in momentum space. Instead of $\eta $, we use the momentum $%
\mathbf{k}$ to label the single-particle states with $\varepsilon
_{k}^{0}=\hbar ^{2}k^{2}/(2m)$. The interaction potential in momentum space
is given by 
\begin{equation}
U(\mathbf{q})=(4\pi /3)\,d^{2}(3\cos ^{2}\theta _{\mathbf{q}}-1)\,,
\label{uq}
\end{equation}%
where $\theta _{\mathbf{q}}$ is the angle between $\mathbf{q}$ and the $z$%
-axis.

From the symmetry of the system, at least for not too strong interaction
strength, we anticipate that pairing only occurs between a particle with
momentum $\mathbf{k}$ and another with momentum $-\mathbf{k}$. In other
words, the ground state has the usual BCS form: 
\begin{equation*}
|gs\rangle =\prod_{\mathbf{k}} \,(u_{\mathbf{k}}+v_{\mathbf{k}}C_{\mathbf{k}%
}^{\dag}C_{-\mathbf{k}}^{\dag}) \,|\mathrm{vacuum} \rangle \,.
\end{equation*}%
Consistent with this ground state, the effective Hamiltonian can be written
as 
\begin{eqnarray}
H_{\mathrm{eff}} =\frac{1}{2} \left[ 
\begin{array}{cc}
C_{\mathbf{k}}^{\dag} & C_{-\mathbf{k}}%
\end{array}%
\right] \! \left[ 
\begin{array}{cc}
\epsilon (\mathbf{k}) & \Delta (\mathbf{k}) \\ 
\Delta ^{\ast }(\mathbf{k}) & -\epsilon (\mathbf{k})%
\end{array}%
\right] \! \left[ 
\begin{array}{c}
C_{\mathbf{k}} \\ 
C_{-\mathbf{k}}^{\dag}%
\end{array}%
\right] +E_0 \,,  \label{Heff}
\end{eqnarray}%
where $E_0= \frac{1}{2}\underset{k}{\sum }[\epsilon (\mathbf{k})-U_{f}(%
\mathbf{k})\langle C_{\mathbf{k}}^{\dag}C_{\mathbf{k}}\rangle -\Delta (%
\mathbf{k})\langle C_{\mathbf{k}}^{\dag}C_{-\mathbf{k}}^{\dag}\rangle ]$, 
\begin{eqnarray}
\epsilon (\mathbf{k}) &=& \varepsilon _{k} + U_f(\mathbf{k}) \,,
\label{self1} \\
U_f(\mathbf{k}) &=& -\underset{\mathbf{k}^{\prime}}{\sum }\, U(\mathbf{k}%
^{\prime}-\mathbf{k})\langle C_{\mathbf{k}^{\prime}}^{\dag}C_{\mathbf{k}%
^{\prime}}\rangle \,, \\
\Delta (\mathbf{k}) &=& \underset{\mathbf{k}^{\prime}}{\sum }\, U(\mathbf{k}-%
\mathbf{k}^{\prime})\langle C_{-\mathbf{k}^{\prime}}C_{\mathbf{k}%
^{\prime}}\rangle \,.  \label{gap1}
\end{eqnarray}
Note that the Hartree term $U_{h}(\mathbf{k})=U(0)\underset{\mathbf{k}%
^{\prime}}{\sum }\langle C_{\mathbf{k}^{\prime}}^{\dag}C_{\mathbf{k}%
^{\prime}}\rangle$ vanishes as, for dipolar itneraction, $U(0)=0$. In
addition, it is easy to see that $U_f(\mathbf{k}) = U_f (-\mathbf{k})$ and $%
\Delta(\mathbf{k}) = -\Delta(-\mathbf{k}) $.

The effective Hamiltonian (\ref{Heff}) takes the diagonalized form 
\begin{equation}
H_{\mathrm{eff}} = \frac{1}{2}\underset{\mathbf{k}}{\sum }\,[E(\mathbf{k}%
)\gamma _{\mathbf{k}}^{\dag}\gamma _{\mathbf{k}}-E(\mathbf{k})\gamma _{-%
\mathbf{k}}\gamma _{-\mathbf{k}}^{\dag}] +E_0 \,,
\end{equation}
in terms of the quasi-particle operators 
\begin{equation*}
\left[ 
\begin{array}{c}
\gamma _{\mathbf{k }} \\ 
\gamma _{\mathbf{k }}^{\dag}%
\end{array}%
\right] =\left[%
\begin{array}{cc}
u_{\mathbf{k}}^* & v_{\mathbf{k}}^* \\ 
-v_{\mathbf{k}} & u_{\mathbf{k}}%
\end{array}
\right] \,\left[ 
\begin{array}{c}
C_{\mathbf{k }} \\ 
C_{\mathbf{k }}^{\dag}%
\end{array}%
\right] \,,
\end{equation*}
with 
\begin{eqnarray*}
u_{\mathbf{k}}^2 = \frac{1}{2} \left(1 + \frac{\epsilon (\mathbf{k})}{E(%
\mathbf{k})} \right) \,, \quad v_{\mathbf{k}}^2 = \frac{1}{2} \left(1 - 
\frac{\epsilon (\mathbf{k})}{E(\mathbf{k})} \right) \,.
\end{eqnarray*}%
where $E(\mathbf{k})=\sqrt{\epsilon (\mathbf{k})^{2}+|\Delta (\mathbf{k}%
)|^{2}}$ represents the quasi-particle dispersion relation.

We remark that the quasi-particle dispersion $E(\mathbf{k})$ may appear to have a similar
form as that in the usual BCS theory for a two-component Fermi
system with contact interaction. There is however a notable difference: In the usual BCS theory,
the Hartree-Fock term is ignored as it represents a constant energy
shift and can be absorbed into the definition of the chemical potential. By
contrast, here the Hartree-Fock contribution (for the uniform system considered
here, only the Fock term survives) is anisotropic, due to the anisotropy of
the dipolar interaction, and must be included explicitly. In fact, even for
quite modest dipolar interaction strength, the Fock term has important
effects and can lead to quite significant deformation of the Fermi surface \cite%
{taka,sogo}.

At thermal equilibrium, we have $\langle \gamma _{\mathbf{k}}^{+}\gamma _{%
\mathbf{k}}\rangle =f(E(\mathbf{k}))$, $\langle \gamma _{-\mathbf{k}}\gamma
_{-\mathbf{k}}^{+}\rangle =1-f(E(\mathbf{k}))$, where $f(x) = 1/(1+e^{\beta
x})$ is the Fermi-Dirac distribution function. Consequently, the self energy
term (\ref{self1}) and the pairing term (\ref{gap1}) take the following
forms: 
\begin{eqnarray}
\epsilon(\mathbf{k})\! \!&=&\! \! \varepsilon _{k} \!- \! \underset{\mathbf{k%
}^{\prime}}{\sum }\,U(\mathbf{k}^{\prime}\! - \! \mathbf{k}) \left[\frac{1}{2%
}-\frac{\epsilon (\mathbf{k}^{\prime})}{2E(\mathbf{k}^{\prime})} \tanh \frac{%
\beta E(\mathbf{k}^{\prime})}{2} \right] ,  \label{self} \\
\Delta(\mathbf{k} )\!\! &=& \! -\underset{\mathbf{k}^{\prime}}{\sum }\,U(%
\mathbf{k}-\mathbf{k}^{\prime})\frac{\Delta (\mathbf{k}^{\prime})}{2E(%
\mathbf{k}^{\prime})}\tanh \frac{\beta E(\mathbf{k}^{\prime})}{2} \,.  \notag
\end{eqnarray}
It is known that the gap equation is ultraviolet divergent. The origin of
the divergence can be attributed to the fact that the dipolar interaction
potential used here [Eq.~(\ref{uq})] is not valid for large momentum. For
large momentum, or equivalently for short distance, the dipolar interaction
potential should be significantly modified due to repulsion between
electrons. Just as in the treatment of two-component Fermi gas with contact
interaction, we need to regularize the interaction in the gap equation. This
problem has been investigated by Baranov and coworkers \cite{baranov}. In
short, the bare dipolar interaction potential $U(\mathbf{k}-\mathbf{k}%
^{\prime})$ in the pairing term should be replaced by the vertex function 
\begin{equation*}
\Gamma(\mathbf{k}-\mathbf{k}^{\prime}) = U(\mathbf{k}-\mathbf{k}^{\prime}) -
\sum_{\mathbf{q}}\, \Gamma(\mathbf{k}-\mathbf{q}) \frac{1}{2\varepsilon_q^0}
U(\mathbf{q}-\mathbf{k}^{\prime}) \,,
\end{equation*}
and the gap equaton should be renormalized as 
\begin{equation}
\Delta(\mathbf{k} )\! = \!- \underset{\mathbf{k}^{\prime}}{\sum }\,\Gamma(%
\mathbf{k}-\mathbf{k}^{\prime})\Delta (\mathbf{k}^{\prime}) \left[\frac{%
\tanh \frac{\beta E(\mathbf{k}^{\prime})}{2} }{2E(\mathbf{k}^{\prime})}- 
\frac{1}{2\varepsilon_{k^{\prime}}^0} \right] \,.  \label{gap}
\end{equation}
Equations (\ref{self}) and (\ref{gap}), together with the number equation $%
N=\sum_{\mathbf{k}} n(\mathbf{k})$ where 
\begin{equation*}
n(\mathbf{k})= |u_{\mathbf{k}}|^2 f(E(\mathbf{k})) +|v_{\mathbf{k}}|^2
(1-f(E(\mathbf{k})) \,,
\end{equation*}
is the momentum distribution function, comprise a complete description of
the dipolar Fermi gas and need to be solved self-consistently.

\begin{figure}[tbp]
\includegraphics[width=8.3cm]{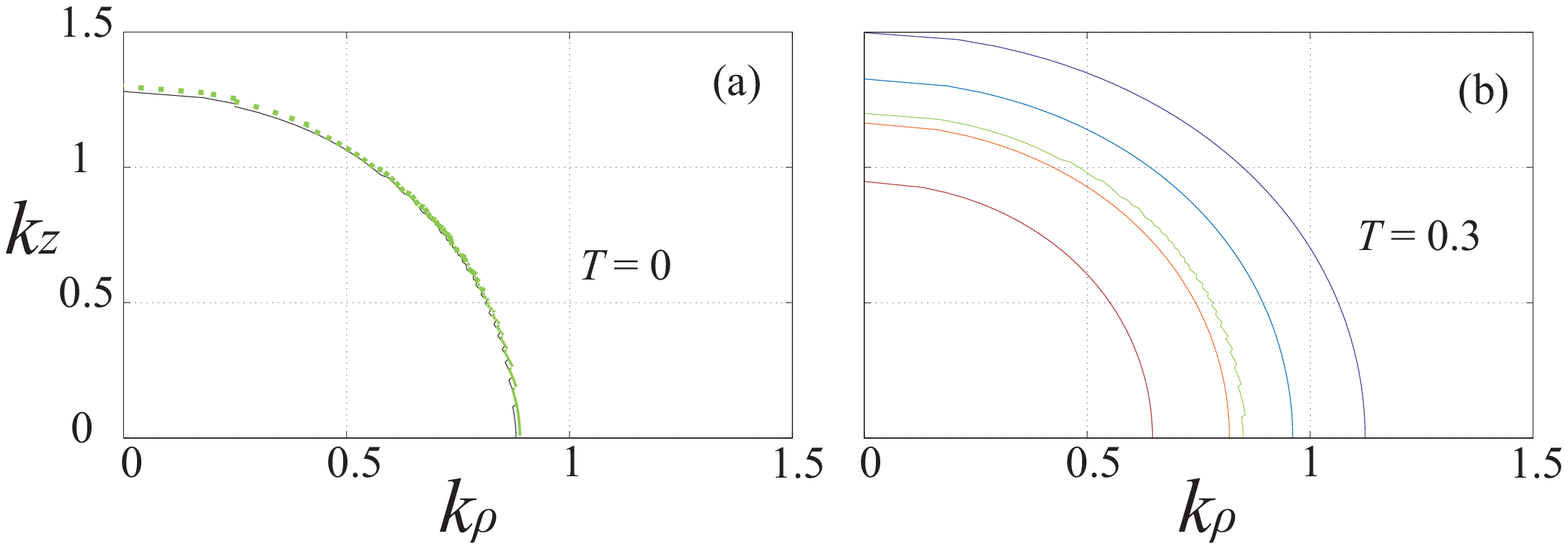}
\caption{(color online) Contour of the momentum distribution function $n(%
\mathbf{k})$ for $C_{dd}=1$ at temperatures $T=0$ (a) and 0.3 $T_F$ (b). In
(a), we draw the Fermi surface. The solid line is the Fermi surface obtained
from the self-consistent calculation of this work, the dotted line is the
one obtained from the variational approach developed in Refs.~\protect\cite%
{taka,sogo}. In (b), the lines from outside to inside correspond to $n(%
\mathbf{k})=0.1$, 0.3, 0.5, 0.7 and 0.9, respectively.}
\label{fig1}
\end{figure}

\emph{Results. ---} Now we present some results. First, let us consider a
normal dipolar gas by taking $\Delta (\mathbf{k})=0$. Note that $\Delta =0$
is always a solution to the gap equation (\ref{gap}). Fig.~\ref{fig1}
ilustrates the momentum distribution function $n(\mathbf{k})$ as a function
of $k_{z}$ and $k_{\rho }=\sqrt{k_{x}^{2}+k_{y}^{2}}$, for two different
temperatures. Here the momentum is in units of the Fermi wave number of the
non-interacting system $k_{F}=(6\pi ^{2}n)^{1/3}$. The dipolar interaction
strength is fixed at $C_{dd}=1$ where $C_{dd}=md^{2}n^{1/3}/\hbar ^{2}$ is
the dimensionless dipolar strength \cite{sogo}. $C_{dd}=1$ corresponds to
the RbK molecule created at the JILA experiment at a modest density of about 
$4\times 10^{-12}$cm$^{-3}$. At $T=0$, $n(\mathbf{k})=v_{\mathbf{k}%
}^{2}=\Theta (-\epsilon (\mathbf{k}))$, where $\Theta (.)$ is the step
function. We draw in Fig.~\ref{fig1}(a) the contour of the Fermi surface. In
Refs.~\cite{taka,sogo}, we developed a variational approach and assume that
the Fermi surface of the dipolar gas has an ellipsoidal shape: 
\begin{equation*}
n(\mathbf{k})=\Theta \left( 1-\alpha ^{2}k_{z}^{2}-k_{\rho }^{2}/\alpha
\right) \,.
\end{equation*}%
where $\alpha $ is the variational parameter characterizing the deformation
of the Fermi surface. At $C_{dd}=1$, we obtain $\alpha =0.7769$. In Fig.~\ref%
{fig1}(a), the dotted line represents the contour of the Fermi surface from
this variational calculation. As one can see, the variational result matches
with the full self-consistent calculation very well. At larger $C_{dd}$,
small difference can be see between the two results. In general, the
variational results exhibit slightly stronger deformation. The same
conclusion has been reached by Ronen and Bohn \cite{ronen}. Fig.~\ref{fig1}%
(b) shows the momentum distribution at $T=0.3$ $T_{F}$, respectively, where $%
T_{F}=E_{F}/k_{B}$ is the Fermi temperature of the non-interacting system.
At finite temperature, Fermi surface gets smeared out. However, the
anisotropy of the momentum distribution is still quite clear.


\begin{figure}[tbp]
\includegraphics[width=8.3cm]{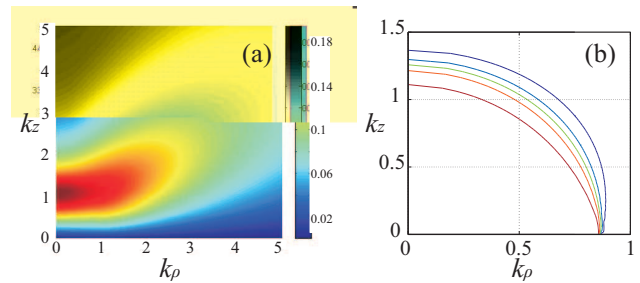}
\caption{(color online) (a) Gap parameter $\Delta(\mathbf{k})$ (in units of $%
E_F$) at $T=0$ for $C_{dd}=1$. (b) The corresponding contour plot of the
momentum distribution function $n(\mathbf{k})$. The lines from outside to
inside correspond to $n(\mathbf{k})=0.1$, 0.3, 0.5, 0.7 and 0.9,
respectively.}
\label{fig3}
\end{figure}
\begin{figure}[ptb]
\begin{center}
\includegraphics[
width=3.3in
]{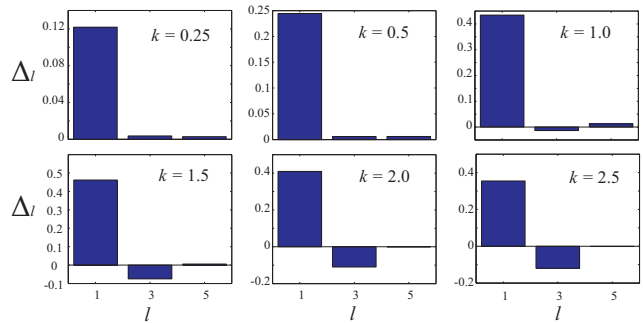}
\end{center}
\caption{$\Delta_l (k)$ for $C_{dd}=1$ at zero temperature.}
\label{fig4}
\end{figure}

Let us now turn to the discussion of the superfluid state. For simplicity,
we take the first-order Born approximation by replacing the vertex function $%
\Gamma(\mathbf{k}-\mathbf{k}^{\prime})$ in the gap equation (\ref{gap}) by
the bare dipolar interaction $U(\mathbf{k}-\mathbf{k}^{\prime})$. This
should be a good approximation as long as the dipolar interaction strength
is not too strong \cite{baranov}. In Fig.~\ref{fig3}(a), we plot the
zero-temperature gap parameter $\Delta(\mathbf{k})$ for $C_{dd}=1$. $\Delta(%
\mathbf{k})$ is an odd function of $\mathbf{k}$ and vanishes for $k_z=0$. As
a consequence, the Fermi surface smears out except at $k_z=0$, as can be
seen from the momentum distribution shown in Fig.~\ref{fig3}(b). The peak
value of $\Delta$ reaches nearly 0.2$E_F$ for this rather modest dipolar
interaction strength, and occurs near $k_z=k_F$ and $k_\rho=0$. To
investigate the angular distribution of $\Delta$, we note that: 
\begin{equation*}
\Delta (\mathbf{k}) = \Delta (k,\cos \theta_{\mathbf{k}}) = \sum_{\mathrm{odd%
}\,l} \Delta_l(k) Y_{l 0}(\cos \theta_{\mathbf{k}})\,,
\end{equation*}%
where $k=|\mathbf{k}|$ and 
due to the cylindrical symmetry of the system, only odd $l$ and $m=0$
components are present. In Fig.~\ref{fig4}, we plot $\Delta_l(k)$ for $%
C_{dd}=1$. For small values of $k$ ($k \lesssim k_F$), $\Delta$ is dominated
by the $l=1$ ($p$-wave) component. For larger $k$, contribution from higher
partial waves may become important.

\begin{figure}[tbp]
\begin{center}
\includegraphics[
width=6cm
]{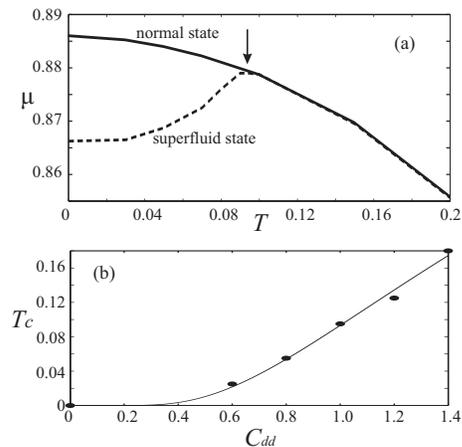}
\end{center}
\caption{(a) Chemical potential $\protect\mu $ as a function of temperature
for superfluid state (dashed line) and normal state (solid line). Energy and
temperature are in units of $E_{F}$ and $T_{F}$, respectively. Here $C_{dd}=1
$. The arrow indicates the location of $T_{c}$. Superfluid state exists for $%
0<T<T_{c}$. (b) $T_{c}$ as a function of $C_{dd}$. Here the dots are
numerical data and the smooth curve is a fit according to Eq.~(\protect\ref%
{fit}).}
\label{fig5}
\end{figure}

Next, we illustrate the finite-temperature effects in Fig.~\ref{fig5}.
The dashed line in Fig.~\ref{fig5}(a) represents the chemical potential of
the superfluid state as a function of temperature. It increases with
temperature. In comparison, the chemical potential of the normal state (the
solid line in Fig.~\ref{fig5}(a)) is a monotonically decreasing function of
temperature. Fig.~\ref{fig5}(b) shows how $T_{c}$ varies with $C_{dd}$. The
solid line is a fit according to 
\begin{equation}
T_{c}/T_{F}=0.8363\,\exp (-2.194/C_{dd})\,.  \label{fit}
\end{equation}%
Some remarks are in order. First, from our calculation, we find that $%
T_{c}\approx 0.1T_{F}$ at $C_{dd}=1$. If we were dealing with a
two-component Fermi gas with contact interaction, such a critical
temperature would correspond to a system inside the unitary regime. As we have
mentioned, $C_{dd}=1$ is a quite modest value for polar molecules.
Therefore, typical polar molecules can easily reach the \textquotedblleft
strongly interacting" regime. Second, it is instructive to compare Eq.~(\ref%
{fit}) to the critical temperature found by Baranov \emph{et al.} \cite%
{baranov} which in our notation takes the form: 
\begin{eqnarray*}
T_{c}/T_{F} &\approx &1.44\,\exp \left[ -\frac{\pi ^{3}}{4(6\pi
^{2})^{1/3}C_{dd}}\right]  \\
&=&1.44\,\exp (-1.9887/C_{dd})\,.
\end{eqnarray*}%
One can notice that the coefficients in the exponent agree quite well. Less
agreement is found in the prefactor. This is, however, understandable as
there are several differences in our treatment. For example, Baranov \emph{%
et al.} have included beyond-mean-field fluctuation and the contribution
from the second-order Born approximation \cite{note}, while neglected the Fock term in
their calculation. 

Finally, to reveal the interplay between pairing and Fermi surface
deformation, we artifically turn off the Fock term in our calculation. We
find that the presence of the Fock exchange interaction increases both the critical
temperature and the magnitude of the order parameter by $20\sim 25\%$. This enhancement can be understood in the following way. The presence
of the Fock term causes an ellipsoidal deformation of the Fermi surface in such a way that it stretches
the momentum distribution along the $z$-axis. As a result, the density of
states near the Fermi surface is increased along $z$ and reduced along the transverse directions. On
the other hand, the dipolar-induced pairing is dominated by the $p$-wave
symmetry, i.e., strongest in the $z$ direction. Therefore, the Fock
interaction-induced Fermi surface deformation tends to enhance superfluid pairing.

\emph{Conclusion. ---} We have presented a fully self-consistent
Hartree-Fock-Bogoliubov theory to study a system of spinless fermions with
long-range interaction. We applied this theory to uniform polar Fermi
molecules, calculated the superfluid order paramter and the critical
temperature $T_{c}$. Our work shows that: a typical Fermi gas of polar
molecules can easily reach the \textquotedblleft strongly
interacting\textquotedblright\ regime with $T_c$ being a
significant fraction of Fermi temperature $T_F$, and the Fock interaction has the effect of
enhancing superfluid pairing. In the future, it will be of great interest to
investigate the collective excitations of the superfluid dipolar Fermi gases, and the effects of quantum fluctuations and the possibility of novel
quantum phases that may arise at large dipolar interaction strength
and/or in the presence of opitical lattice potential \cite{lattice}.

This work is supported by the NSF, the Welch Foundation (Grant No. C-1669)
and the W.M. Keck Foundation. HP acknowledges the hospitality of Aspen Center for Physics where part of the work is completed.


\begin{thebibliography}{99}
\bibitem{note}Quantum fluctuations tend to decrease $T_c$, while the second-order Born approximation tends to increase $T_c$ \cite{baranov}. Hence these two effects tend to partially cancel each other.
\end{thebibliography}
\end{document}